\documentclass[twocolumn,prx,superscriptaddress,aps]{revtex4}

\usepackage{ulem}
\usepackage{bbm}
\usepackage{color}
\usepackage{amsmath}
\usepackage{graphicx}
\usepackage{hyperref}
\usepackage{comment}
\usepackage{siunitx} 
\usepackage{verbatim}
\usepackage{subfigure}
\usepackage{amssymb}

\definecolor{mygreen}{rgb}{0,0.5,0}
\definecolor{mygrey}{rgb}{0.5,0.5,0.5}
\definecolor{myred}{rgb}{0.75,0,0}
\definecolor{myblue}{rgb}{0,0,0.75}
\definecolor{mymagenta}{cmyk}{0,1,0,0.12}
\definecolor{mycyan}{cmyk}{1,0,0,0.12}
\definecolor{myorange}{rgb}{1,0.5,0}
\definecolor{myviolet}{rgb}{0.5,0.0,0.75}

\newcommand{\commentout}[1]{}

\newcommand{\agg}[1]{\hat{#1}^{\dag}}
\newcommand{\media}[1]{\langle #1  \rangle}

\newcommand{\Ea}[1]{\agg{a}_{\scriptscriptstyle #1}}

\newcommand{\ket}[1]{\ensuremath{\left|#1\right\rangle}}

\newcommand{\CESPDC}{{CE-SPDC}}
\newcommand{\Lxtal}{{L}}

\newcommand{\lockfreq}{\nu_{\text{lock}}}
\newcommand{\reffreq}{\nu_{\text{ref}}}
\newcommand{\laserfreq}{\nu_{\text{laser}}}
\newcommand{\sigfreq}{\nu_{s}}
\newcommand{\idlfreq}{\nu_{i}}
\newcommand{\pumpfreq}{\nu_{p}}
\newcommand{\seedfreq}{\nu_{\text{seed}}}
\newcommand{\aomifreq}{\nu_{\text{AOM}1}}
\newcommand{\aomiifreq}{\nu_{\text{AOM}2}}

\graphicspath{{../Images/}}

\begin{document}

\newcommand{\mytitle}{Narrowband photon pairs with independent frequency tuning for quantum light-matter interactions.}

\title{\mytitle}

\newcommand{\ICFO}{ICFO - Institut de Ciencies Fotoniques, The Barcelona Institute of Science and Technology, 08860 Castelldefels, Barcelona, Spain}
\newcommand{\ICREA}{ICREA - Instituci\'{o} Catalana de Recerca i Estudis Avan{\c{c}}ats, 08010 Barcelona, Spain}
\newcommand{\ZH}{State Key Laboratory of Modern Optical Instrumentation, College of Optical Science and Engineering, Zhejiang University, Hangzhou 310027,China}

\author{Vindhiya Prakash}
\affiliation{\ICFO}
\author{Lorena C. Bianchet}
\affiliation{\ICFO}
\author{Marc T. Cuairan}
\affiliation{\ICFO}
\author{Pau Gomez}
\affiliation{\ICFO}
\author{Natalia Bruno}
\affiliation{\ICFO}
\author{Morgan W. Mitchell}
\affiliation{\ICFO}
\affiliation{\ICREA}




\begin{abstract}
We describe a cavity-enhanced spontaneous parametric down-conversion (CE-SPDC) source for narrowband photon pairs with filters such that over 97$\%$ of the correlated photons are in a single mode of \SI{4.3+-0.4}{\mega\hertz} bandwidth. Type-II phase matching, a tuneable-birefringence resonator, MHz-resolution pump tuning, and tuneable Fabry-Perot filters are used to achieve independent signal and idler tuning. We map the CE-SPDC spectrum using difference frequency generation to precisely locate the emission clusters, and demonstrate CE-SPDC driven atomic spectroscopy. The generated photon pairs efficiently interact with neutral rubidium, a well-developed system for quantum networking and quantum simulation. The techniques are readily extensible to other material systems.

\end{abstract}

\maketitle

{Light-matter quantum interface technologies are built upon the coherent interaction of non-classical light with material quantum systems \cite{PredojevicBook2015}.  Several suitable material systems have been developed, including single atoms \cite{SchlosserN2001}, ions \cite{LeuchsJMO2013} and molecules \cite{WriggeNP2007}, as well as ensembles of atoms \cite{SparkesNJP2013}, ions \cite{ClausenNJP2013}, and impurities in solids \cite{DeRiedmattenN2008}. Such light-matter interfaces employ resonant or near-resonant optical response of the material to achieve a strong interaction with propagating \cite{TeyNJP2009} or cavity-bound \cite{HennrichPRL2005} photons. The participating photons must be matched to the optical transitions of the material system, requiring a control of spatial mode, frequency and bandwidth, with negligible contamination by photons of unwanted frequencies. The workhorse method for tailoring the properties of photon pairs is cavity-enhanced (CE) spontaneous parametric downconversion (SPDC) \cite{OuPRL1999}. In \CESPDC, high-frequency ``pump'' photons from a laser spontaneously decay to produce time-correlated pairs of lower-frequency ``signal'' and ``idler'' photons in a nonlinear crystal. The crystal is contained within an optical resonator that both enhances the spectral brightness via the Purcell effect \cite{PurcellPR1946}, and determines the possible spatial modes, frequencies and bandwidth of emission. }

{\CESPDC~ has proven to be a reliable technique to generate bright \cite{Weinfurter2000} and narrowband photons for interaction with matter \cite{ScholzAPL2007, Wolfgramm1, JWPan_2008, WangOC2010, AWhite_2016, Benson_2016}. Several works have used heralded single photons from \CESPDC~to excite single trapped atoms, quantum dots and solid state quantum memories \cite{MWM_2009, SchuckPRA2010, PiroNP2011, ClausenN2011, ZhangNP2011, RielanderPRL2014, Eschner_2015, Marquardt_2015, Paudel_2019, tsai2019}, while a few have used photon pairs in resonant interaction with a single system \cite{WolfgrammNPhot2013}, and even with two distinct systems via optical frequency conversion \cite{FeketePRL2013, SeriPRX2017}. All these material-resonant sources could tune one of the output photons, sometimes over a large range\cite{Marquardt_2015}, with a consequent effect on the other photon's frequency. To date, however, no \CESPDC~source has been demonstrated with independent tuning of the \CESPDC~signal and idler. Though this is not required for applications that employ one of the photons as a herald, there are tasks in quantum light-matter interactions that require this capability.  For example, driving multi-photon processes in a single material system, or direct interaction with distinct systems. Most material systems of interest have multiple, closely-spaced internal levels due to, e.g., hyperfine splitting, and multi-photon processes among these internal levels, e.g. stimulated Raman transitions, involve photons of different frequencies. Many of them also experience environment-induced energy shifts, e.g. crystal field shifts in solids or trap-induced level shifts in vacuum. To address specific transitions in these, a general-purpose \CESPDC~source should have independent tuning of the frequencies of signal and idler, derived from an absolute frequency reference, e.g. an atomic or molecular spectroscopic feature.  Here we describe such a source, designed for resonant excitation of optically-trapped atomic rubidium, a well-developed system for strong light-matter interactions in free space \cite{SchlosserN2001, TeyNJP2009, KaufmanS2014, LabuhnN2016, JenneweinPRL2016, ChinNC2017, Bruno:19}.  Using a combination of time-correlated photon counting and atomic spectroscopic methods, we demonstrate a single mode output and independent tuneability over the D1 line, a preferred line for slow and stopped light, quantum memories, and proposals for atom-mediated photon-photon interactions. While the \CESPDC~apparatus is designed to produce photon pairs that can interact with this single system, our techniques can be readily applied to other wavelengths for interaction with other material systems or combinations of them.}

\begin{figure*}[t]
\centering
\includegraphics[scale=0.7]{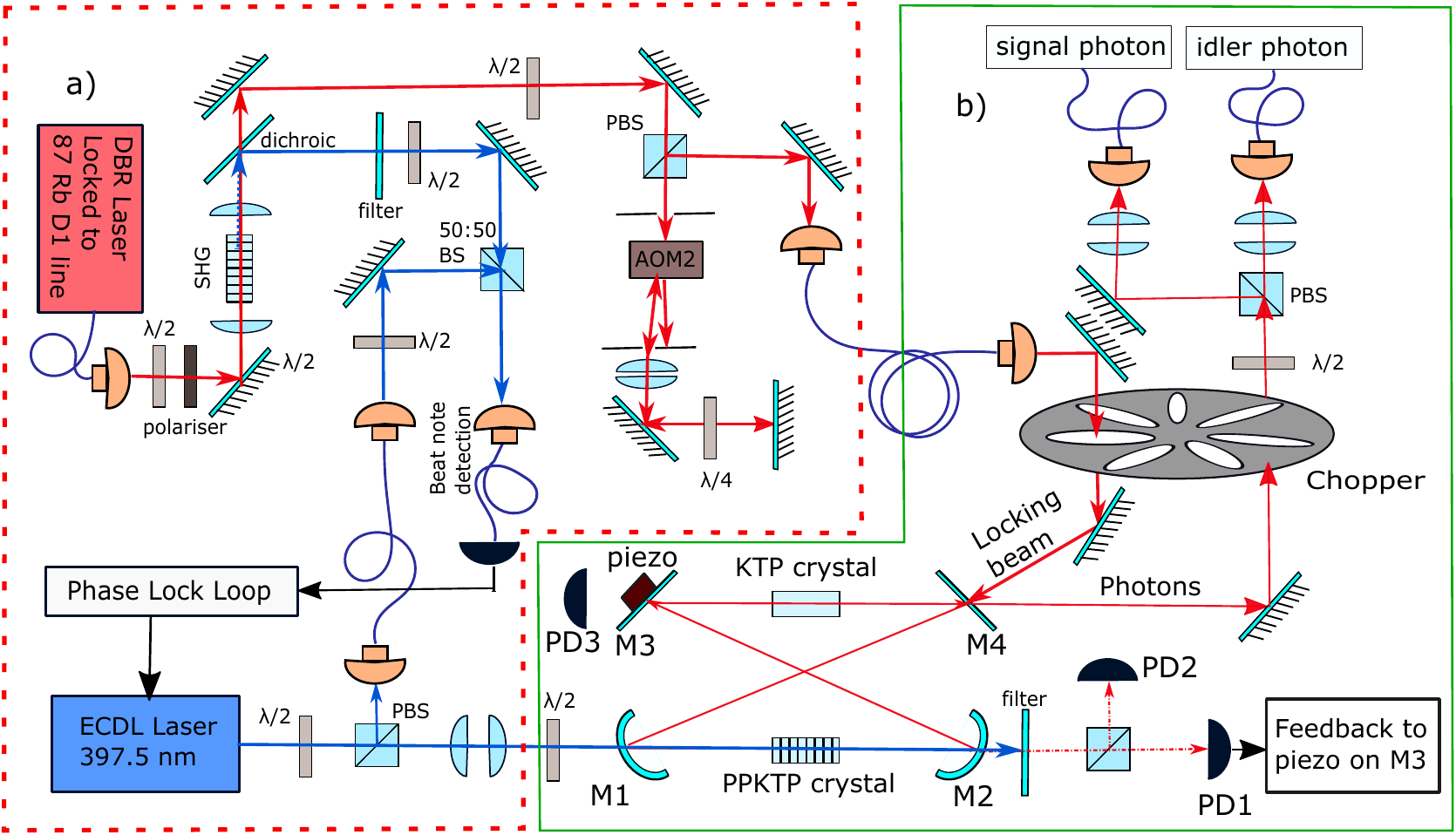}
\caption[Schematic of Optics] { {Schematic of lasers and \CESPDC~source.  a) Laser systems.  A distributed Bragg reflector at \SI{795}{\nano\meter}, locked to a transition of the $^{87}$Rb D1 line, is upconverted by second harmonic generation (SHG).  The undepleted \SI{795}{\nano\meter} light is frequency shifted with an acousto-optic modulator (AOM) and used to lock the \CESPDC~cavity.  An external-cavity diode laser at \SI{397}{\nano\meter} is stabilized relative to the \SI{795}{\nano\meter} second harmonic by a beat-note lock.  b) Narrowband pair source, consisting of a bow-tie cavity containing an SPDC crystal (PPKTP), pumped by the ECDL and a tuning crystal (KTP).  The locking beam enters the cavity through the out-coupling mirror (M4) and co-propagates with pump, signal and idler.  Transmission of the locking beam through mirror M2 is used to stabilize the cavity length. The chopper blocks the locking beam when the photons are collected. Signal and idler photons are split based on polarization. }  PBS: polarizing beamsplitter, $\lambda/2$: half-wave plate, $\lambda/4$: quarter-wave plate,  PD: photodetector. }
\label{fig: Optics Schematic} 
\end{figure*}

Such a source opens the way for studies of resonant multi-photon effects at the most fundamental, individual-quantum level.  One major motivation for such studies are the many proposals for strong photon-photon interactions  \cite{HarrisPRL1998, HarrisPRL1999, ChangNP2007, HuangN2009, KolchinPRL2011, PeyronelN2012, PerczelPRL2017}, of interest to quantum information processing.
To date these mechanisms have only been tested with classical light \cite{MitchellPRA2000, ChenS2013,VolzNP2014}.

The article is organized as follows: In \autoref{sec:Expt Setup} we describe the CE-SPDC source. In \autoref{sec:SpectralContent} we theoretically model the CE-SPDC spectrum and use difference frequency generation to measure the model parameters. In \autoref{sec:FilterCavity} we describe the design and construction of a tunable filter capable of efficiently transmitting a single CE-SPDC mode while blocking all other modes. In \autoref{sec:characterisation} we present measurements to characterize the resulting downconverted photon pairs and demonstrate their independent tuneability by rubidium vapour absorption spectroscopy.

\section{Experimental Setup}\label{sec:Expt Setup}

To produce independently-tunable photon pairs with frequencies and linewidths suitable for interaction with the Rb D1 line, we constructed a \CESPDC~source, with auxiliary Fabry-Perot (FP) filters.

\subsection{The Cavity Enhanced SPDC source}

The experimental set-up is shown schematically in  \autoref{fig: Optics Schematic}. 
The \CESPDC~ optical resonator is in a bow-tie configuration of four mirrors (\autoref{fig: Optics Schematic}B), with the two closest to the SPDC crystal being concave, to produce a cavity mode with a \SI{30}{\micro\meter}  beam waist  inside the SPDC crystal.  Mirrors M1 to M3 are coated for \SI{99.9}{\percent} reflectivity at \SI{795}{\nano\meter} whereas the outcoupling mirror M4 has nominal reflectivity of \SI{97}{\percent} at this wavelength.  Scattering losses from the super-polished, ion-beam sputtered mirrors are negligible, and the weak transmission of M1, M2 and M3 is useful for locking and probing the cavity resonances. These four mirrors form a resonator for the  \SI{795}{\nano\meter} signal and idler photons, whereas the \SI{397}{\nano\meter} pump light is not resonated because the same mirror coatings are largely transparent for this wavelength. See Wolfgramm \textit{et al.} \cite{Wolfgramm1} for more details of the cavity construction. The SPDC occurs in a \SI{20}{\milli\meter} long periodically-poled potassium titanyl phosphate (PPKTP) crystal (henceforth called SPDC crystal), with a poling period of \SI{9.4}{\micro\meter} to enable  Type-II phase matching from vertically-polarized ($V$) pump to vertically-polarized signal and horizontally-polarized ($H$) idler beams. The pump is focused into the SPDC crystal with a beam waist of \SI{30}{\micro\meter}, exits through M2 and is blocked. Throughtout this article we use subscripts $p$, $s$ and $i$, to refer to physical quantities pertaining to the pump, signal and idler respectively. The FWHM of the crystal phase-matching efficiency was calculated to be 148 GHz \cite{wolfgrammthesis} at this wavelength.  The cavity length can be adjusted via a piezo-electric actuator on mirror M3.  The cavity also contains an unpoled KTP crystal (tuning crystal) with the same dimensions as the PPKTP crystal and with the optical axis aligned parallel to that of the PPKTP crystal. The birefringence of the tuning crystal alters the round-trip optical path length of the signal relative to the idler and thereby provides a degree of freedom for independently tuning the signal and idler cavity resonances. Both the SPDC and tuning crystals are mounted in PTFE (teflon) ovens with optical access. The crystals' temperatures are independently controlled using Peltier elements and\commentout{ \SI{5}{\kilo\ohm}} NTC thermistors, with the controllers being an analog PID\commentout{(Wavelength Electronics HTC1500)} with its set point fixed by a microcontroller.\commentout{(Arduino Uno + Analog Shield).} This technique gives a temperature resolution of $\approx$ \SI{5}{\milli\kelvin} and a long-term stability of \SI{1}{\milli\kelvin}.

The effective cavity length is 610 mm. To make a classical measurement of the cavity linewidth and finesse, we perform cavity transmission spectroscopy. We scan  the frequency of a \SI{795}{\nano\meter} distributed Bragg reflector (DBR) laser, which enters the cavity via M4 and is collected behind M2. We simultaneously collect Rb D1 saturated absorption spectra as an absolute frequency reference.  We observe transmission resonances of linewidth \SI{8.8\pm0.4}{\mega\hertz}, which, when adjusted for the $\SI{1.2\pm0.5}{\mega\hertz}$ short-term linewidth of the DBR laser, implies a cavity (power) decay rate of $\gamma = 2 \pi \times \SI{7.6\pm0.6}{\mega\hertz}$. We note that the signal or idler FWHM linewidth emitted by SPDC with a single-frequency pump and equal decay rates for signal and idler is $\gamma \,(2^{1/2}-1)^{1/2}\approx 0.64 \gamma$ \cite{OBenson3}. Thus a photon from our source is expected to have a linewidth of \SI{4.9\pm0.4}{\mega\hertz}, similar to the \SI{5.75}{\mega\hertz} natural linewidth of the D1 transition, $5^{2}\text{S}_{1/2}$ to $5^{2}\text{P}_{1/2}$ in Rb. The Free Spectral Range (${\rm FSR}$) of the signal and idler are slightly different due to the birefringence induced difference in optical path length between them. The same cavity spectroscopy technique also allows us to measure ${\rm FSR}_s$ and ${\rm FSR}_i$, from which we compute ${\rm FSR}_{\rm mean} \equiv ({\rm FSR}_s + {\rm FSR}_i)/2  = \SI{496\pm8}{\mega\hertz}$. The observed $\gamma$ and ${\rm FSR}_{\rm mean}$ values are consistent with the cavity finesse (${\cal F}$) of 66 predicted from the measured outcoupler transmission and the reflection losses on the faces of the crystal \cite{wolfgrammthesis}. The difference $\Delta{\rm FSR} \equiv {\rm FSR}_s - {\rm FSR}_i = \SI{3.5\pm0.5}{\mega\hertz}$ can be more precisely estimated by difference frequency generation (DFG), as described in \autoref{sec:DFG}.

\subsection{ Frequency stabilisation scheme for tuneable photon pairs} \label{sec:tunablephotons}
\begin{figure}[t]
\begin{center}
\includegraphics[width=0.95\columnwidth]{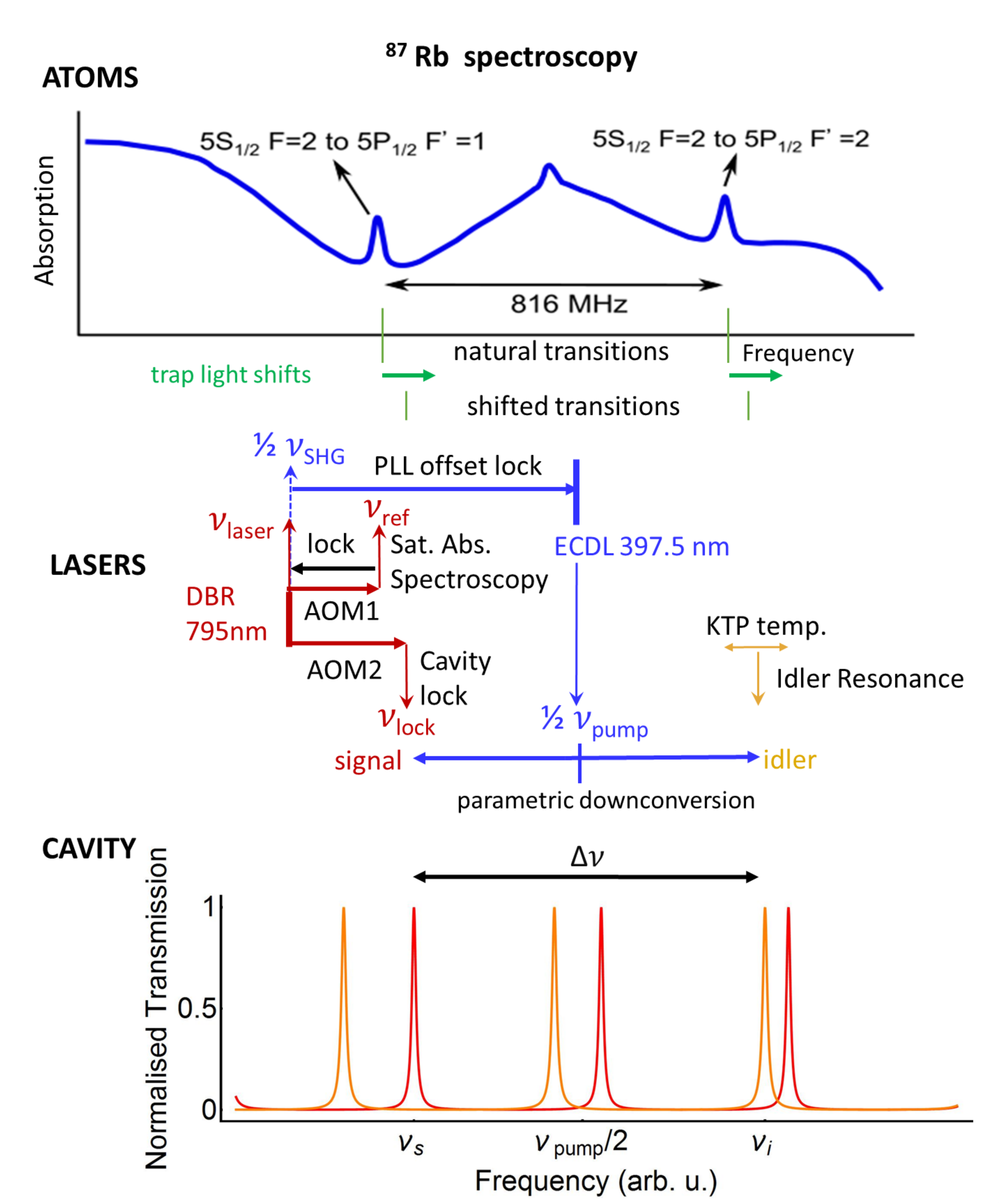}
\caption{Illustration of frequencies employed in the \CESPDC~ source.  Scenario shown achieves the configuration: $\sigfreq$ tuned to the (light-shifted) $F=2 \rightarrow F'=1$ transition and $\idlfreq$ tuned to the (light-shifted) $F=2 \rightarrow F'=2$ transition of the $^{87}\text{Rb}$ D1 line. Top graph (``atoms'') shows the saturated absorption spectrum (in blue) with light-shifted transitions shown below the horizontal axis in green.   Middle section (``lasers'') shows frequency relationships among frequencies described in the text.  Frequency separations are not to scale.  Lower section (``cavity'') shows cavity spectrum including signal (red) and idler (orange) modes.  The $\Delta{\rm FSR}$ and $\gamma$ are exaggerated for clarity.
}
\label{fig:TuneabilityScheme}
\end{center}
\end{figure}

A schematic of the frequency control setup is shown in \autoref{fig: Optics Schematic}A) and a diagrammatic representation of the relevant frequencies is shown in \autoref{fig:TuneabilityScheme}. Light from a \SI{795}{\nano\meter} DBR laser diode is  double-passed through an acousto-optic modulator (AOM) driven by an amplified voltage-controlled oscillator (VCO), and stabilized to a desired transition in the $^{87}$Rb D1 spectrum at frequency $\reffreq$ using saturated absorption spectroscopy. The laser beam, at frequency $\laserfreq = \reffreq - \aomifreq $  is upconverted to $\nu_{\text{SHG}}$, equal to $2\laserfreq$, by second-harmonic generation (SHG) in a PPKTP crystal in a single pass configuration. The upconverted light is separated using a dichroic mirror and the unconverted \SI{795}{\nano\meter} light is double-passed through a second AOM, thereby shifting its frequency to $\lockfreq = \laserfreq + \aomiifreq$ and adding a weak frequency modulation at 2.7 MHz to enable a dither lock of the SPDC cavity. The tuning range of the AOMs allows $\aomifreq$ and  $\aomiifreq$ to be independently set in the range from \SI{156}{\mega\hertz} to \SI{164}{\mega\hertz}. Thus $\lockfreq$ can be detuned from the atomic transition frequency $\reffreq$ by $\aomiifreq - \aomifreq$, to account for light induced frequency shifts in the trapped cold atom(s). This locking light is passed through a spinning-blade mechanical chopper and matched in spatial mode and polarization to the $V$ polarised signal mode of the cavity. A small fraction of this light exits the cavity by transmission through M2 and is collected on PD1, the signal of which is demodulated to obtain an error signal and fed back by a proportional-integral controller to the piezo on M3, thereby stabilizing the signal mode of the SPDC cavity to $\lockfreq$. The chopper blocks the cavity-detector path whenever the laser-cavity path is open, preventing locking photons from reaching the detectors. 
 
With the signal mode stabilized to $\sigfreq = \lockfreq$, the idler mode is also stabilized, at frequency $\idlfreq = \sigfreq + \Delta \nu$, where the offset $\Delta \nu$ is determined by the birefringence of the SPDC and tuning crystals in the cavity.  $\Delta \nu$  is controlled via the temperature of the tuning crystal. The SPDC crystal's temperature is maintained at a value that gives efficient phase-matching, as described in \autoref{sec:SpectralContentAnalysis}. 

A grating stabilised external-cavity diode laser (ECDL) with a central wavelength at 397.5 nm \commentout{(DLC DL PRO HP from TOPTICA Photonics) }is used as the SPDC pump.   A portion of the light from this laser is mode-matched to the upconverted light at $2 \laserfreq$ produced by SHG, and detected using an amplified silicon detector with a bandwidth of 1.5 GHz.\commentout{(FPD310-V from Menlo Systems)} The resulting beat note is stabilized by feeding the detector signal to a radio-frequency digital phase-locked loop (PLL) (ADF4111 from Analog Devices), and using the output of the PLL to stabilise the frequency of the pump laser.\commentout{ in an evaluation board EV-ADF411XSD1Z).}\commentout{fed back to the laser current via the laser controller and locked using the inbuilt DLC Pro side-of-fringe lock.} The resulting frequency lock allows us to lock the pump laser frequency within the range
$\SI{80}{\mega\hertz} 
\leqslant 
\mid \nu_{\text{pump}}-2 \laserfreq \mid \leqslant \SI{1.5}{\giga\hertz}$. The pump frequency is chosen to achieve the energy conservation condition $\sigfreq + \idlfreq = \pumpfreq$.  

With this scheme, we can generate degenerate or non-degenerate photons at \SI{795}{\nano\meter} up to a maximum frequency difference between signal and idler photons of 1.18 GHz, limited by the speed of the beat-note detector for the blue light.

\section{Spectral content of the \CESPDC~output}\label{sec:SpectralContent}

While the set-up described this far generates narrowband photons at the required frequencies, its output also contains a significant fraction of photons in other, unwanted modes. We theoretically model and experimentally analyse the spectral content of the \CESPDC~ as follows to design a filter capable of isolating a single emission mode. 

\subsection{Theory}\label{sec:theory}

The two-photon state of the \CESPDC~ output can be written as,
\cite{MandelSPDCTheory, WalmsleySPDCQuantumDesc}
\begin{equation}
\ket{\psi}= \int d\omega_s\int d\omega_i  \: \phi (\omega_s,\omega_i)\, \Ea{s}(\omega_s)\Ea{i}(\omega_i)\ket{0}.
\end{equation}
where $\Ea{s}(\omega_s)$ and $\Ea{i}(\omega_i)$ are creation operators for the respective frequencies, $\phi (\omega_s,\omega_i)$ is the joint spectral function (JSF) and $|0\rangle$ indicates the vacuum state.  The joint spectral intensity (JSI), $|\phi (\omega_s,\omega_i)|^2$,  is simply derived using the ``unfolded cavity'' approach described by Jeronimo-Moreno \textit{et al.} \cite{jeronimo2010theory}. For a single-frequency pump, this gives
\begin{equation}\label{eq:JSIcavity}
|\phi (\omega_s,\omega_i)|^2 \propto 
\delta(\omega_p- \omega_i-\omega_s)\:
\text{sinc}^2\bigg(\frac{\Delta k \Lxtal }{2}\bigg)\: |A_{s}(\omega_{s})|^2 |A_{i}(\omega_{i})|^2,
\end{equation}
where $\Delta k$ is the wave-vector mismatch in the quasi phase matched PPKTP crystal of length $\Lxtal$ and 

\begin{equation}\label{eq:ASquared}
|A_{\epsilon}(\omega)|^2 \propto \left[ 1+ \left( \frac{2 \mathcal{F}}{\pi}\right) ^2 \text{sin}^2 \left( \frac{\omega}{2 \text{FSR}_{\epsilon}}\right) \right]^{-1},
\end{equation}
where $\mathcal{F}$ is the cavity finesse and $\epsilon = s\text{ or }i$. The FSR is the  separation of adjacent longitudinal modes in linear frequency, given by ${\rm FSR}_\epsilon = 
{c}[\sum_x n_{x}^{(g)}({\bf p}_{\epsilon}, \omega_{\epsilon}, T_x) L_x]^{-1}$, where the sum is over the different regions (SPDC crystal, tuning crystal, and air spaces) traversed in a single round-trip, ${\bf p}$ indicates the polarization, $L_x$ and $T_x$ the length and temperature of each region respectively.  $n^{(g)} = n + \omega dn/d\omega$ indicates the group index, which we compute using tabulated Sellmeier coefficients from \cite{Kato}.

The JSI of \autoref{eq:JSIcavity} is the product of four factors of differing selectivity: the $\delta$-function describes exact energy conservation in the SPDC process, the ${\rm sinc}^2$ function describes the efficiency of phase-matching, which for our crystal and wavelength has a FWHM of $\approx \SI{150}{\giga\hertz}$, and the two $|A(\omega)|^2$ functions each describe a comb of peaks, separated by the respective FSRs. The finesse of the signal and idler, which in principle can be different, are in fact the same for our source as we operate close to degeneracy and there is no polarisation-dependent loss in the cavity.  Because the signal and idler ${\rm FSR}$s are each $\approx \SI{0.5}{\giga\hertz}$, roughly 300 modes of both signal and idler can be independently contained within the phase-matching bandwidth. However, only pairs of signal-idler modes that satisfy energy conservation contribute to the JSI and of these, those that do so to within a cavity linewidth will be strongest in the JSI.

Unless specific compensation measures are taken \cite{Wolfgramm1}, a type-II SPDC cavity will  have ${\rm FSR}_s \neq \text{FSR}_i$, due to birefringence of the nonlinear media.  This FSR mismatch provides a means to restrict the modes into which photon pairs are generated. \commentout{ For example, if a signal frequency $\nu_s^{(0)}$ and its energy-conserving partner $\nu_i^{(0)} = \nu_p - \nu_s^{(0)}$ are both perfectly resonant, i.e., if  $\nu_s^{(0)} = l \, {\rm FSR}_s$ and $\nu_i^{(0)} = m \,{\rm FSR}_i$ for some integers $l$ and $m$, then the frequency of the neighboring signal mode $\nu_s^{(1)} =  \nu_s^{(0)} +  {\rm FSR}_s$ will be resonant, but its energy-conserving partner $\nu_i^{(1)} =  \nu_i^{(0)} - {\rm FSR}_s$ will not be perfectly resonant as the resonance will be centered at $ \nu_i^{(0)} - {\rm FSR}_i$}. While it is possible to avoid emission on neighboring modes by  making the FSR mismatch, $\Delta{\rm FSR}$, large compared to the linewidth \cite{Geneva1}, this is not the case in our cavity. If $|\Delta \text{FSR}| \lesssim \Delta \nu$, there will be significant emission into at least a few mode pairs close to any ideally-matched pair. This gives rise to the \textit{clustering effect} \cite{Eckardt, Luo_2015, Geneva1, Henderson}. The output of the \CESPDC~is in clusters of modes which repeat with a frequency spacing given by \cite[see Eq.~(25)]{Eckardt} 
\begin{equation}\label{eq:clusterspacing}
\Delta\nu_{\text{cluster}}=\frac{\text{FSR}_s \: \: \text{FSR}_i}{|\Delta{\rm FSR}|}.
\end{equation}
If there exists an energy-conserving mode pair $\nu_s^{(l)} = l \, {\rm FSR}_s$ and $\nu_i^{(m)} = m \, {\rm FSR}_i$ for some $l,n\,\in \mathbbm{Z}$, it will be the brightest in the cluster. Recalling that $\gamma/(2 \pi) = ({\rm FSR}_s + {\rm FSR}_i)/(2 {\cal F})$ is the linewidth of signal and idler,  the modes $\nu_s^{(l+M)}$ and $\nu_i^{(m-M)}$, for $M \in \mathbbm{Z}$, would have half the brightness of $\nu_s^{(l)}$ and $\nu_i^{(m)}$ if $M\, \Delta \text{FSR}=\pm {\gamma}/({4 \pi})$. 
The same applies to the modes  $\nu_i^{(l-{M})}$and $\nu_i^{(m+{M})}$. Thus we can expect the total number of modes within the FWHM of a cluster to be  
\begin{equation}\label{eq:modespercluster}
2\,M+1=\frac{\gamma}{4\pi |\Delta{\rm FSR}| }=\frac{{\rm FSR}_{\rm mean}}{ \mathcal{F} |\Delta{\rm FSR}|}.
\end{equation} 

\subsection{Observation of Spectral Content through Difference Frequency Generation} \label{sec:DFG} 
\begin{figure}[b]
\centering
\includegraphics[width=0.85\columnwidth]{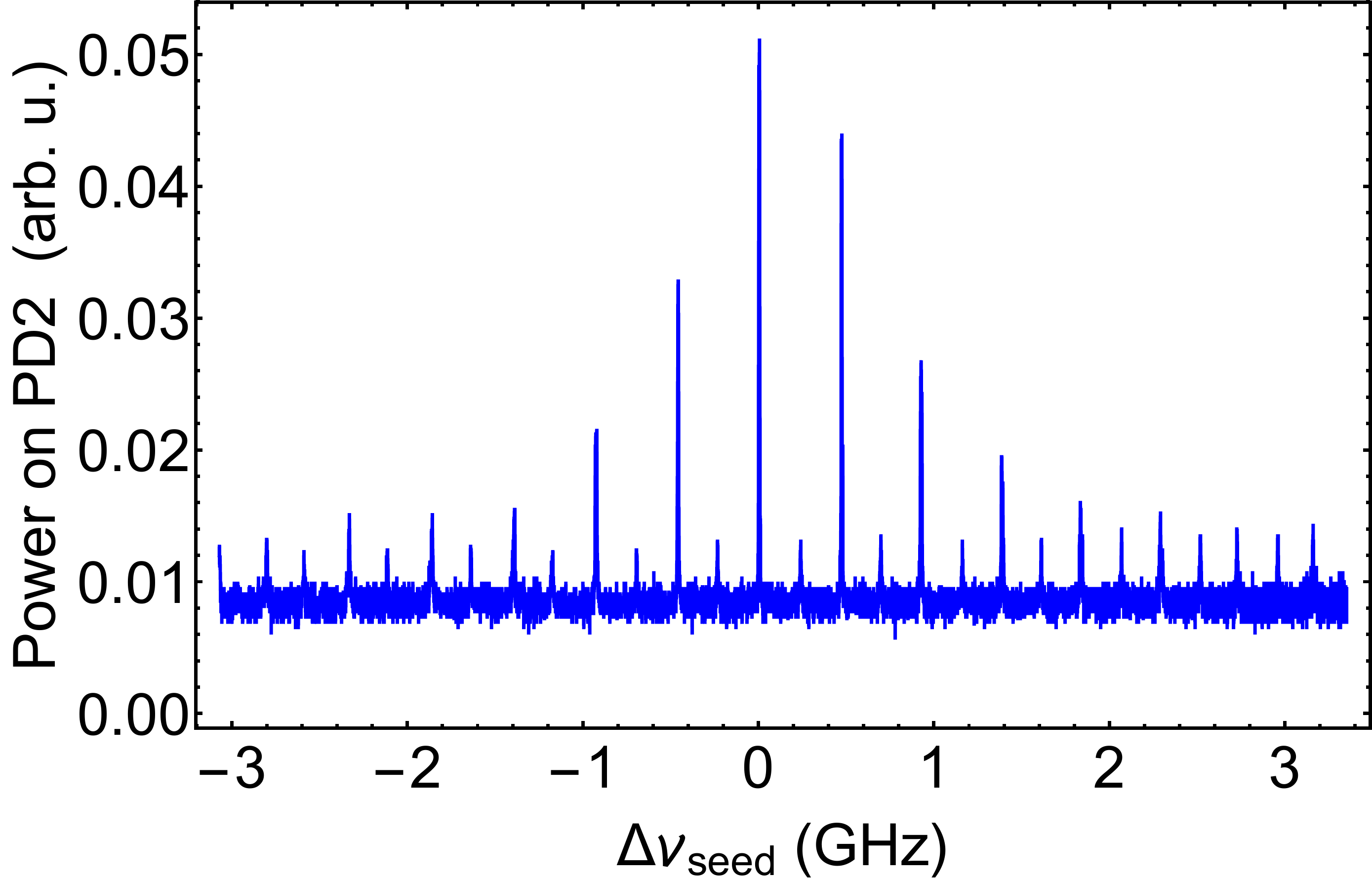}
\caption{Measurement of one cluster of \CESPDC~emission by DFG.  Graph shows generated idler power ($P_i$) as a function of the change in the input seed frequency ($\Delta \seedfreq$) for fixed cavity length and pump frequency. The brightest peak at $\Delta\seedfreq=0$ is due to the simultaneous resonance of $\seedfreq$ and $\idlfreq = \pumpfreq - \seedfreq$ whereas other peaks at $\Delta \seedfreq = \pm \, {\rm FSR_s}, \pm 2 \, {\rm FSR_s}, \ldots $ have decreasing brightness according to the mismatch in resonance with the corresponding idler modes at $\pm \, {\rm FSR_i}, \pm 2 \, {\rm FSR_i}, \ldots $
A second set of peaks, intermediate between DFG peaks and of roughly constant amplitude, appear to be due to a small coupling of the seed beam to a higher transverse mode, and are unrelated to DFG. Background level of $P_i \approx 0.008$ is due to imperfect blocking of the pump light.
}
\label{fig:DFG}
\end{figure}

Using the Sellmeier coefficients for KTP and the measured FSR of the cavity, it is possible to compute all the locations of the \CESPDC~clusters using \autoref{eq:JSIcavity}.  Such predictions are, however, reasonably sensitive to the refractive index of the crystals, which may vary in function of the crystal growth methods.  Rather than rely upon such predictions, we used the {stimulated} parametric down-conversion process, i.e. difference frequency generation (DFG), to measure the parametric gain of the \CESPDC~ \cite{eckstein2014}.  

A $V$ polarized locking beam at frequency $\lockfreq$ (resonant to $F=2$ to $F'=1$ transition in $^{87}$Rb D1 line) was, exclusively for this experiment, made to enter  the cavity in the reverse direction, i.e., opposite to the direction of the pump, signal and idler, and detected in transmission using PD3. The chopper was not used for this experiment. The PPKTP crystal was pumped with \SI{20}{\milli\watt} at $2 \lockfreq - \SI{120}{\mega\hertz}$. About \SI{40}{\milli\watt} of \SI{795}{\nano\meter} light, matched in spatial mode, polarization and direction to the signal mode, was introduced through M4, as a seed for the DFG process.  Light exiting the cavity through M2 was split according to polarization and detected at PD1 and PD2.  The seed frequency $\seedfreq$ was scanned, and DFG was observed as production of $H$-polarised idler light.

Fig \ref{fig:DFG} shows the detected idler power seen on PD2, as  $\seedfreq$ was scanned by $\Delta\seedfreq$. The output shows a cluster of modes that contained about 3 to 4 modes within the FWHM of each cluster. The first cluster was observed when $\seedfreq$ was scanned around \SI{377099.1+-0.5}{\giga\hertz}, as measured with a wavemeter. The cluster repeated itself with a period of \SI{70.7+-0.5}{\giga\hertz} in $\seedfreq$. Using \autoref{eq:clusterspacing} and taking FSR$_{\text{mean}}$ = \SI{496}{\mega\hertz}, we conclude that a cluster spacing of \SI{70.7}{\giga\hertz} occurs when
 $|\Delta {\rm FSR}|$ = \SI{3.5+-0.1}{\mega\hertz}. For modelling the JSI produced by the \CESPDC~ we took FSR$_{\text{mean}}$= \SI{496}{\mega\hertz} and $\Delta$FSR = 3.5 MHz and $\gamma=2\pi \times \SI{7}{\mega\hertz}$.
 
\begin{figure}[b]
\centering
\subfigure[\ ]{
\includegraphics[width=0.77\columnwidth]{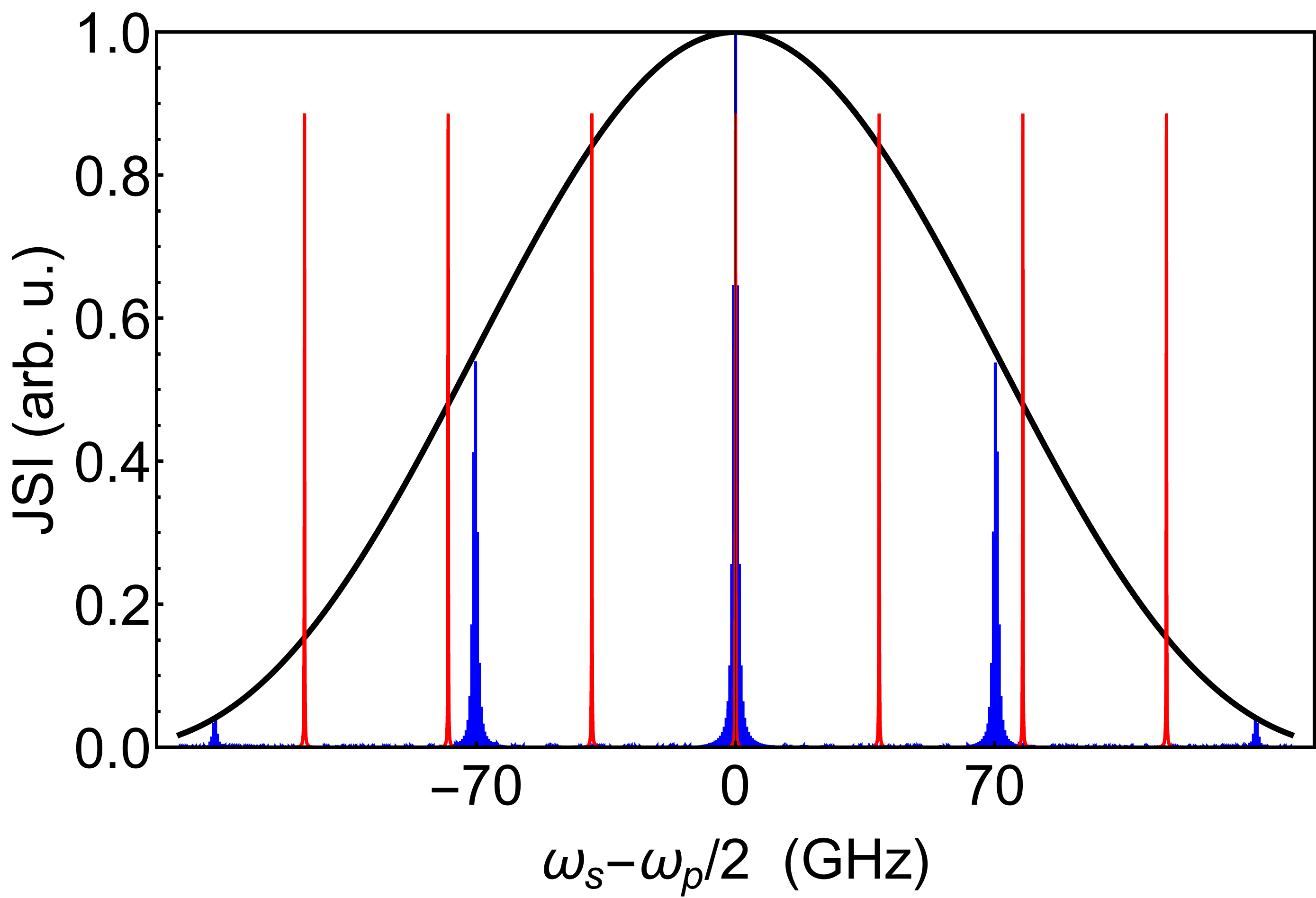}}
\subfigure[\ ]{
\includegraphics[width=0.77\columnwidth]{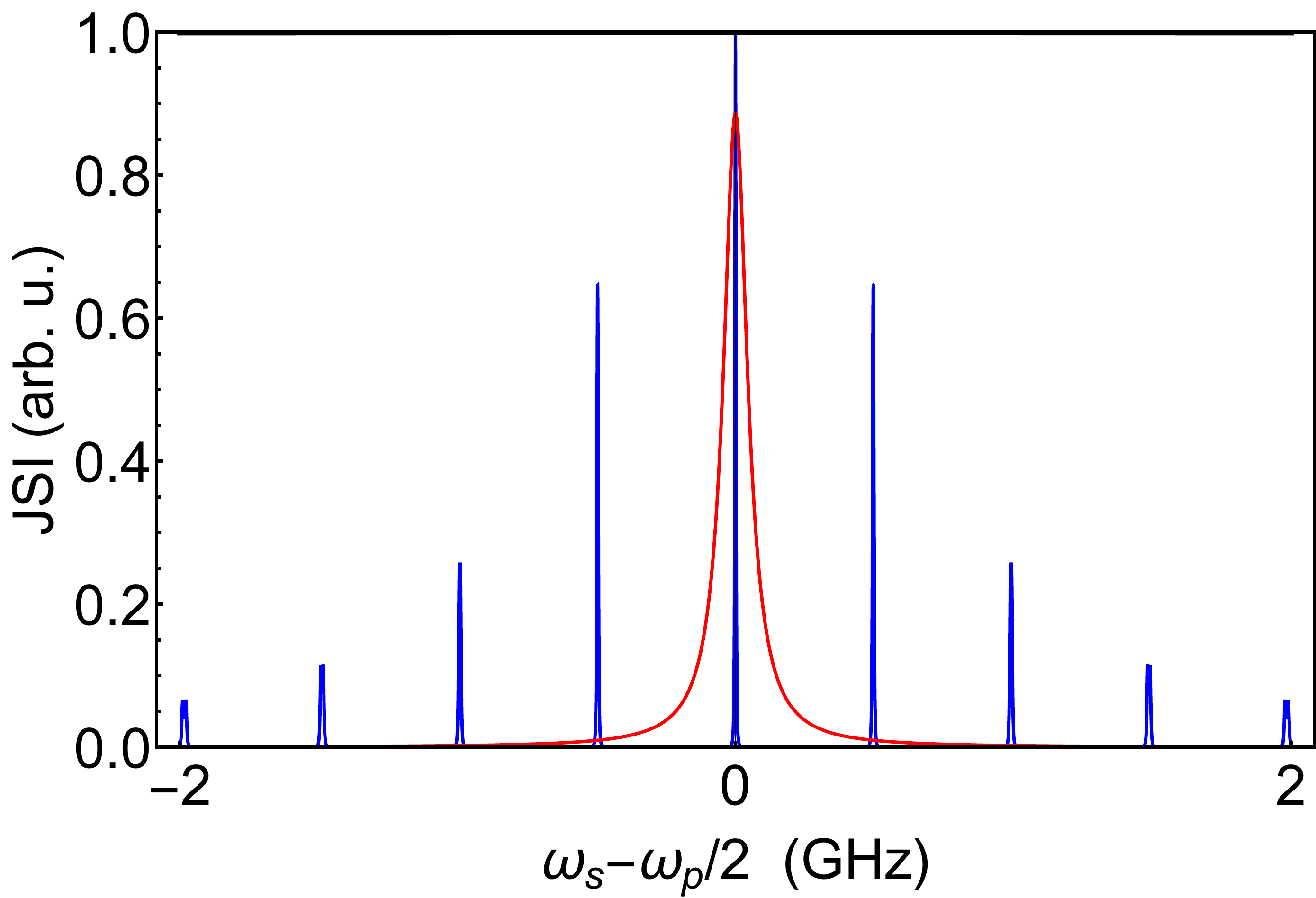}}

\caption{Theoretical two photon JSI from \CESPDC. The blue solid lines of unequal height show the function $|\phi(\omega_s,\omega_p-\omega_s)|^2$ as a function of $\omega_s$ for a constant $\omega_p$. The black thick line shows the phase matching efficiency with the crystal tuned such that the degenerate modes, i.e., $\omega_i=\omega_s=\omega_p/2$, are brightest. The red lines of equal height show $|T_s(\omega_s)|^2$, the transmission of the FP filter. Graphs are plotted for the measured parameters of the bow-tie cavity, FP cavity and phase-matching bandwidth described in the text. Top graph shows the several clusters allowed by phase matching, with the FP cavity set to pass only the central cluster. Bottom graph shows closeup of the central cluster, with the FP cavity set to pass only the central line.}
\label{fig:clusters}
\end{figure}

\section{Filter Cavity}  \label{sec:FilterCavity}
\begin{figure*}[t]
\centering
\includegraphics[scale=0.23]{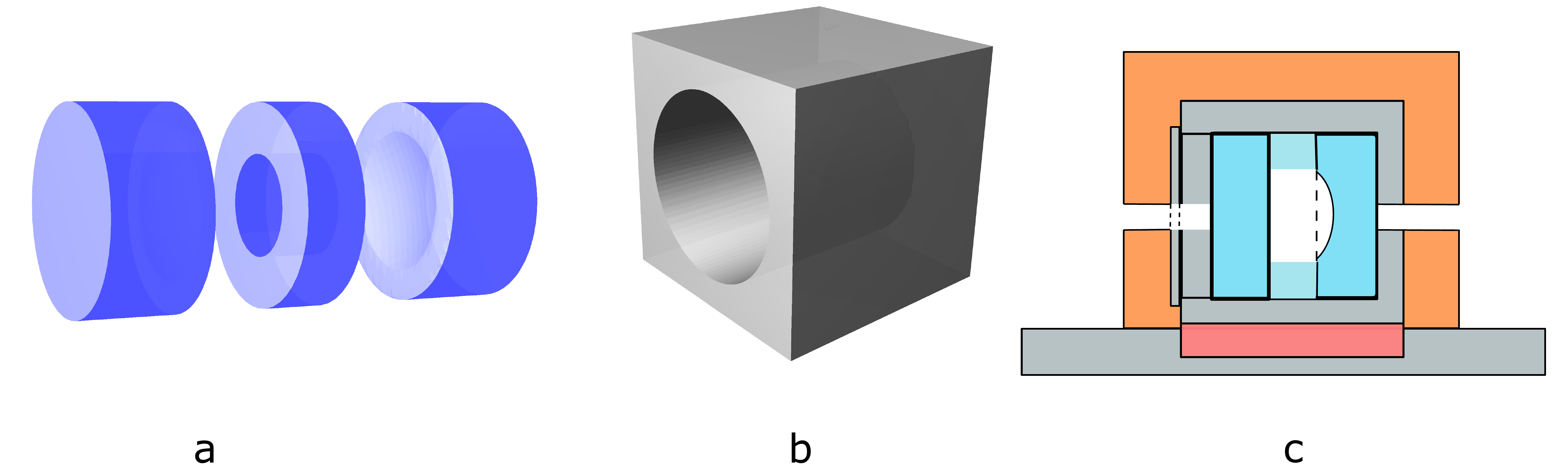}
\caption{FP filter assembly. a) The FP filter consisting of one concave mirror, an annular spacer, and a plano mirror in face-to-face contact is housed in a b) hollowed aluminum block and cemented around the edges with epoxy. c) Vertical cross section of the   filter assembly: aluminum box and heat sink are shown in grey, the Peltier element in red, the mirrors and spacer in shades of blue and the insulator in orange.}
\label{fig:FPcavity}
\end{figure*}

\autoref{fig:clusters} is obtained by computing the JSI using \autoref{eq:JSIcavity} and the parameters estimated from the DFG. As shown in the figure, the \CESPDC~ source is predicted to produce a multi-mode output, emitting on three clusters of $\approx 4$ signal modes, with correlated emission of idler frequencies.  Now having an accurate picture of the spectral content of the \CESPDC~ source, it is possible to design an FP filter that passes one of these signal modes with efficiency approaching unity, while blocking the rest by a large factor.

The JSI of the filtered \CESPDC~ source is 
 \begin{equation}\label{eq:JSIWithFilters}
|\phi_{\rm filt}(\omega_s, \omega_i)|^2=|\phi(\omega_s, \omega_i)|^2 |T_s(\omega_s)|^2   |T_i(\omega_i)|^2
\end{equation}
where the filter transmission function is 
\begin{equation}\label{eq:TSquared}
|T(\omega)|^2 \approx T_{\text{max}}^2 \bigg[ 1+ \bigg( \frac{2 \mathcal{F}_{\rm FP}}{\pi}\bigg)^2 \text{sin}^2 \bigg( \frac{\omega}{2 \text{FSR}_{\rm FP}}\bigg) \bigg]^{-1},
\end{equation}
and the constant $T_{\rm max}^2 \approx 1$ for a low-loss cavity and subscript FP refers to physical quantities for the filter. The filter transmission can be tailored by proper choice of FSR$_{\rm FP}$ and $\mathcal{F}_{\rm FP}$ to discriminate between modes within a cluster, and also to discriminate against clusters other than the central cluster.

\begin{figure}[b]
\centering
\subfigure[\ ]{
\includegraphics[width=0.77\columnwidth]{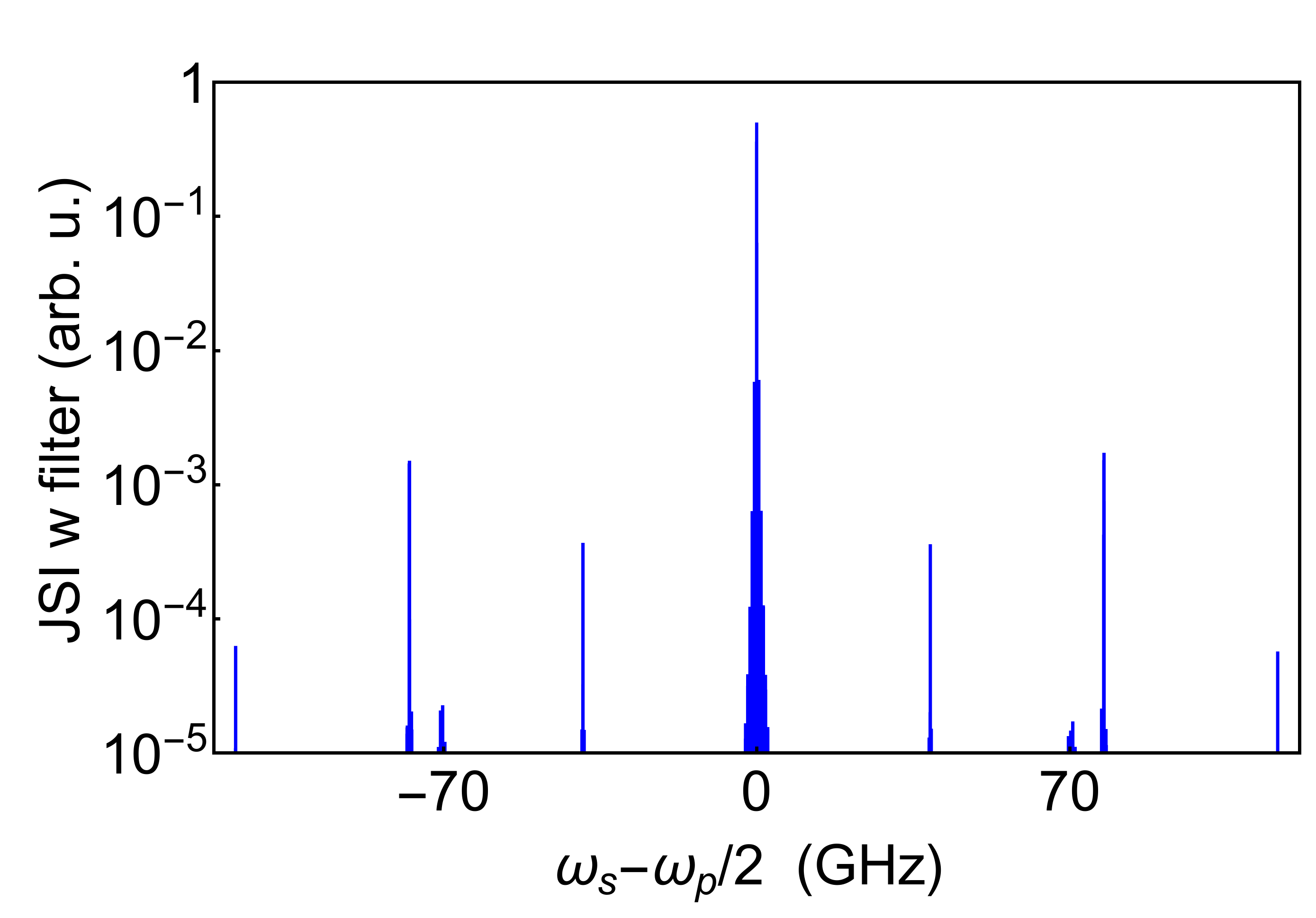}}
\subfigure[\ ]{
\includegraphics[width=0.77\columnwidth]{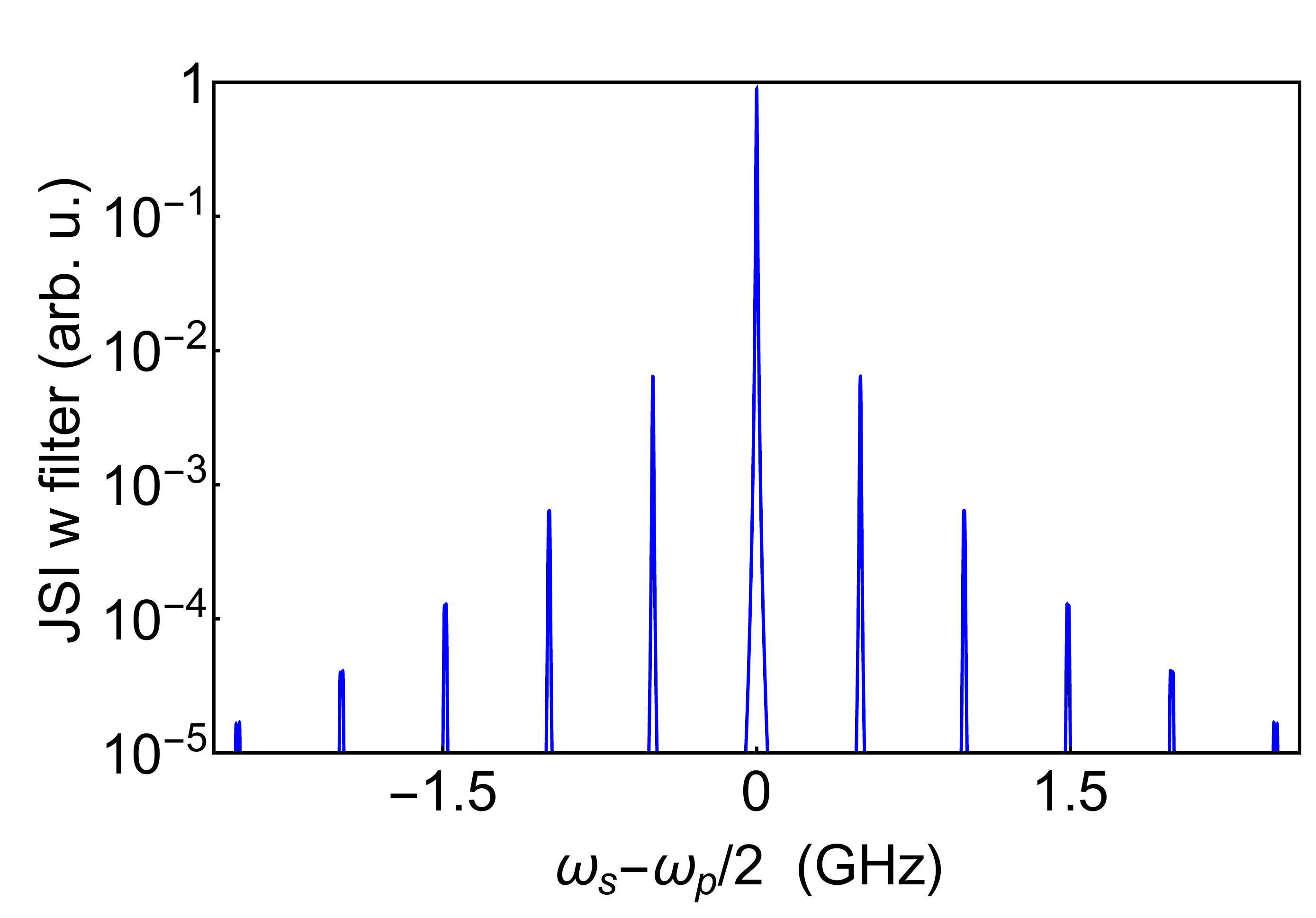}}

\caption{Theoretical two photon JSI from \CESPDC~after the FP filter.  Graphs as in \autoref{fig:clusters}, but $|\phi_{\text{filt}}(\omega_s, \omega_p-\omega_s)|^2$ is shown. This describes the JSI when the filter, described in subsection 3.3, is placed in the signal arm and no filter in the idler arm (so that $|T_i|$ = 1 ).  The contribution of unwanted photons is 2.3 $\%$ within a window \SI{+-1}{\nano\meter} when the filter is set to transmit the brightest mode from the \CESPDC.}
\label{fig:FilterExtinction}
\end{figure}

We now describe the construction of a stable, temperature-tunable Fabry-Perot filter cavity suitable for selecting a single signal or idler mode from the multimode \CESPDC~ output.  The design, shown in Fig.~\ref{fig:FPcavity}, has some similarities to previous filters for this purpose \cite{palittapongarnpim, BensonFilter}, but employs a three-element -- rather than monolithic -- design to more easily achieve the desired linewidth, FSR, and tuning properties.  The cavity is made by one plane mirror (flatness $\lambda /20$) and one concave mirror with a radius of curvature of \SI{-1000}{\milli\meter}.\commentout{(LaserOptik mirror substrate part numbers S-00018 and S-00139).} Both mirrors are of fused silica and have low loss ion beam sputtered multi-layer dielectric mirror coatings on the interior-facing surfaces, with nominal reflectivities $99.2\% \pm 0.1\%$ at \SI{795}{\nano\meter}. The mirrors are anti-reflection (AR) coated on the exterior-facing surfaces. A suitable separation of the two mirrors is achieved with a borofloat annular spacer of \SI{3.8+-0.1}{\milli\meter} thickness with a \SI{5.5}{\milli\meter} diameter hole,\commentout{ (custom made by LaserOptik),} which gives the cavity an FSR$_{\rm FP}$ of \SI{39.4}{\giga\hertz}. The linewidth was measured to be \SI{96.6 +- 0.9}{\mega\hertz} and the on resonance transmission to be 87$\%$. As shown in   \autoref{fig:FilterExtinction}, the discrimination achieved by the filter would be greater than \SI{20}{dB} for all non-selected modes. If such a signal photon is used as a herald, it indicates the presence of an idler photon that is similarly single mode.

The TEM00 mode in the cavity has a beam waist at the plane mirror, with a spot size of \SI{125}{\micro\meter}.  To assemble the filter, the mirror-spacer-mirror stack is first held in face-to-face physical contact, and then a two-component epoxy\commentout{ (Varian Torr Seal)} is applied around the edges to seal the trapped air space and provide structural rigidity. The resulting filter cavity is highly insensitive to vibration and pressure fluctuations. Finesse and transmission measurements after one year of use show no signs of degradation. The filter cavity is housed in a custom-built oven, constructed of an aluminum block with a circular bore to accept the glued mirror assembly, lined with a thermal interface pad \commentout{ (part number : EYG-S091210DP, from Panasonic electronic components),}and with threaded aluminum end caps to hold the assembly in place. The aluminum block is insulated with a covering of \SI{3}{\centi\meter} thick extruded polystyrene and glued with thermally conducting epoxy to a Peltier element which in turn is glued to an aluminum heat dissipator. An\commentout{\SI{10}{\kilo\ohm}} NTC thermistor embedded in the aluminum block and the Peltier element, are used to control the oven temperature, which is stabilised the same way as the crystals in the \CESPDC~.\commentout{  The temperature controller is again a HTC1500 controlled by an Arduino Uno + Analog Shield. The Arduino output voltage resolution of \SI{150}{\micro\volt} corresponds to a The resolution \SI{5}{\milli\kelvin} change in temperature of the sensor at room temperature.} The nonlinearity of the NTC response results in higher resolution at lower temperatures. The resolution is \SI{2.5}{\milli\kelvin} at \SI{15}{\degreeCelsius}, \SI{5}{\milli\kelvin} at room temperature and \SI{9}{\milli\kelvin} at \SI{40}{\degreeCelsius}.

The resonance of the filter can be shifted by 1 FSR$_{\rm FP}$ by changing its temperature by \SI{31.7}{\kelvin}, while a \SI{5}{\milli\kelvin} change corresponds to a shift in resonance of 6 MHz, based on the values for the coefficients of thermal expansion viz., $\SI{3.2e-6}{\per\kelvin}$ and $\SI{5.1e-7}{\per\kelvin}$ respectively for borofloat and fused silica at room temperature. Even in the worst case scenario of having a \SI{10}{\milli\kelvin} temperature resolution, the filter's line-center can be tuned to within 6 MHz of a desired $\sigfreq$ or $\idlfreq.$ This worst-case mismatch implies $< 1\%$ loss of transmission relative to exact resonance. 

\section{Characterization of  the \CESPDC~output}\label{sec:characterisation}

\subsection{Second-order Cross-Correlation Function }
\begin{figure}[b]
\centering
\subfigure[\ ]{
\includegraphics[width=0.77\columnwidth]{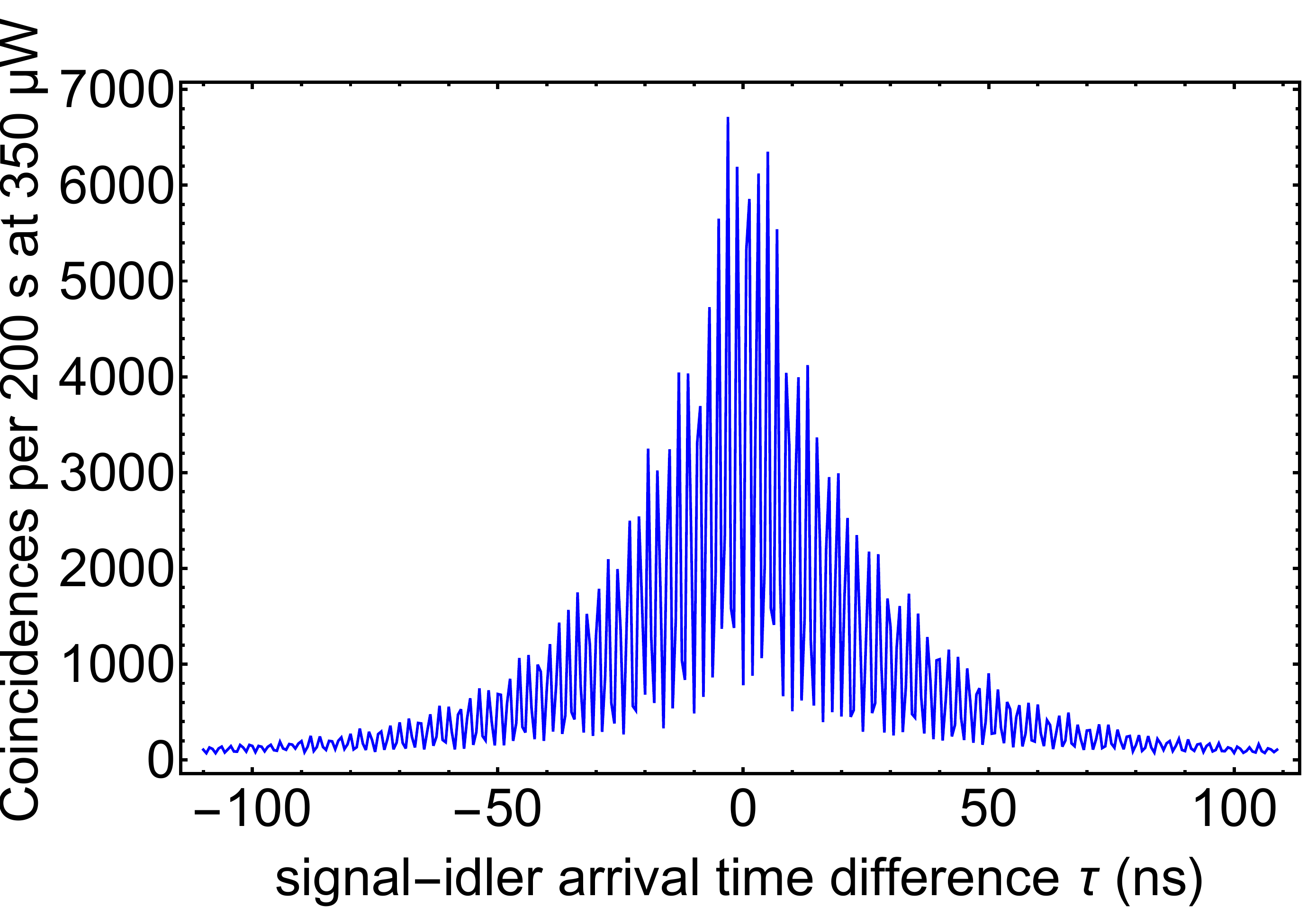}}
\subfigure[\ ]{
\includegraphics[width=0.77\columnwidth]{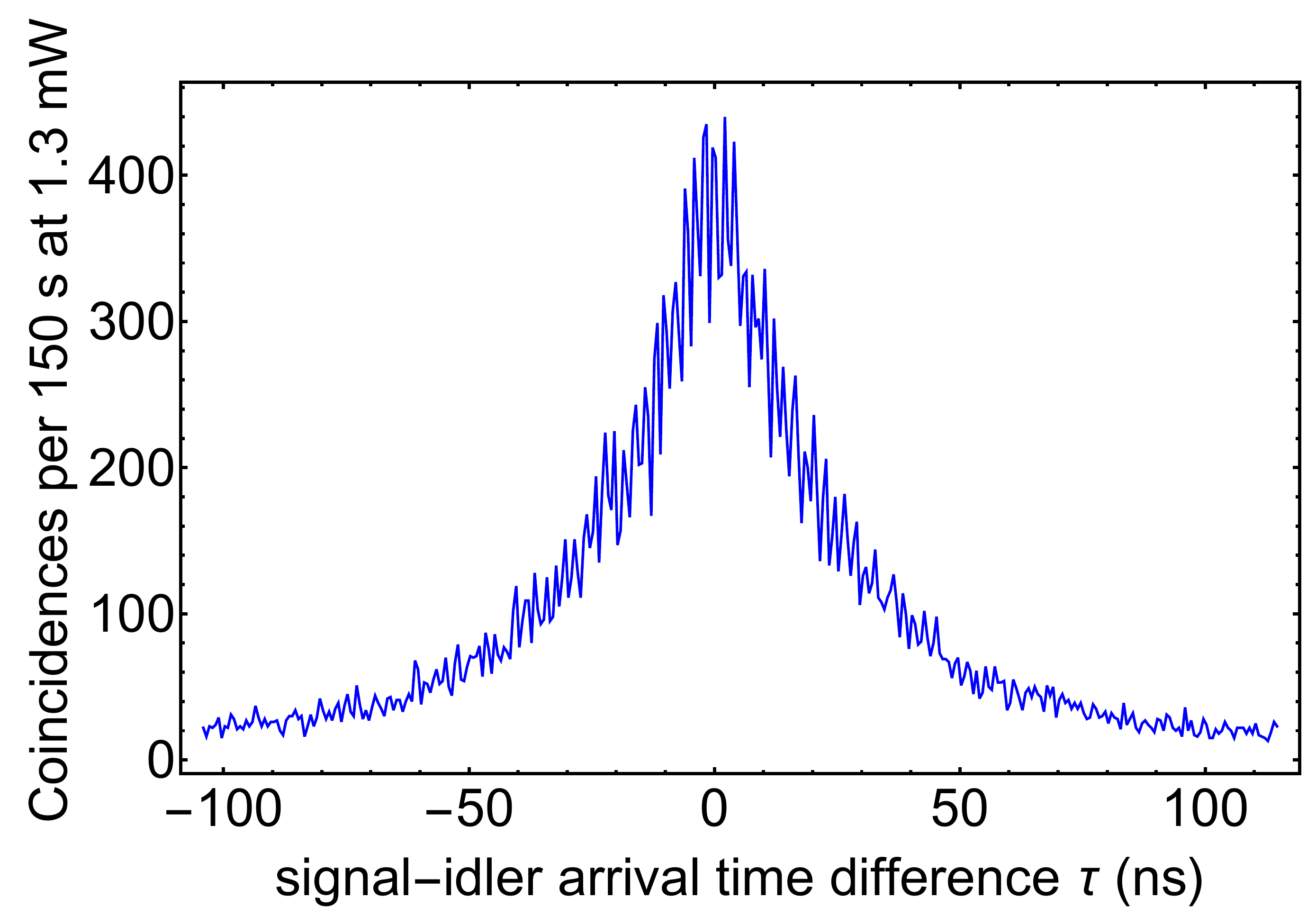}}
\caption{Measurement of the signal-idler cross correlation. We plot the un-normalised cross correlation function $G_{s,i}^{(2)}(\tau)$. a)  Signal-idler coincidence time distribution  from the \CESPDC~ source. The interference between the multimode components results in a comb of peaks separated by the cavity round-trip time $T_{\rm cav} \approx \SI{2}{\nano\second}$, clearly resolved by the $1/\SI{625}{\pico\second}$ TDC time resolution. b) The same distribution measured on \CESPDC~ output with FP filter on the signal channel.  Absence of oscillations indicates absence of emission on modes spaced by less than $1/\SI{625}{\pico\second} = \SI{1.6}{\giga\hertz}$.}
\label{fig:G2}
\end{figure}

We can obtain information about the mode structure of the \CESPDC~ output from the normalized second-order cross-correlation function,
\begin{eqnarray}
g_{s,i}^{(2)}(\tau) & \equiv & \frac{\media{E_s^\dagger(t+\tau)E_i^\dagger(t)E_i(t)E_s(t+\tau)}}
{\media{E_s^\dagger(t)E_s(t)}  \media{E_i^\dagger(t+\tau)E_i(t+\tau)}},
\end{eqnarray}
where $E_\epsilon(t)$ is the positive frequency part of the quantized field for mode $\epsilon$ and $\media{\cdot}$ indicates the average over time $t$. $g_{s,i}^{(2)}(t_s - t_i)$ describes the pair correlations distribution: it is proportional to the probability density for detecting a signal photon at time $t_s$, conditioned on detecting an idler photon at time $t_i$.  As described in \cite{OBenson3}\cite{OBenson2, wolfgramm2, FeketePRL2013}, for a type-II \CESPDC~ source with resonant signal and idler, $g_{s,i}^{(2)}(\tau)$ has the form of a comb of peaks spaced by the cavity round-trip time $T_{\rm cav}$, modulated by a double-exponential envelope. However, when only a single mode of signal-idler pair is incident on the detector, the comb structure - a consequence of multimode intereference- vanishes, leaving only the double exponential envelope to define the $g_{s,i}^{(2)}(\tau)$ \cite{OBenson3}:
\begin{eqnarray}\label{eq:g2}
g_{s,i}^{(2)}(\tau) & \propto  \left\{  
\begin{array}{lr}
\exp[-\frac{1}{2}\gamma_s \tau] & \tau > 0 \\
\exp[\frac{1}{2} \gamma_i \tau] & \tau < 0 
\end{array} 
\right.,
\end{eqnarray}
where $\gamma_s$ and $\gamma_i$ are the cavity damping rates    for signal and idler, respectively. 

\begin{figure*}[t]
\centering
\includegraphics[scale=0.38]{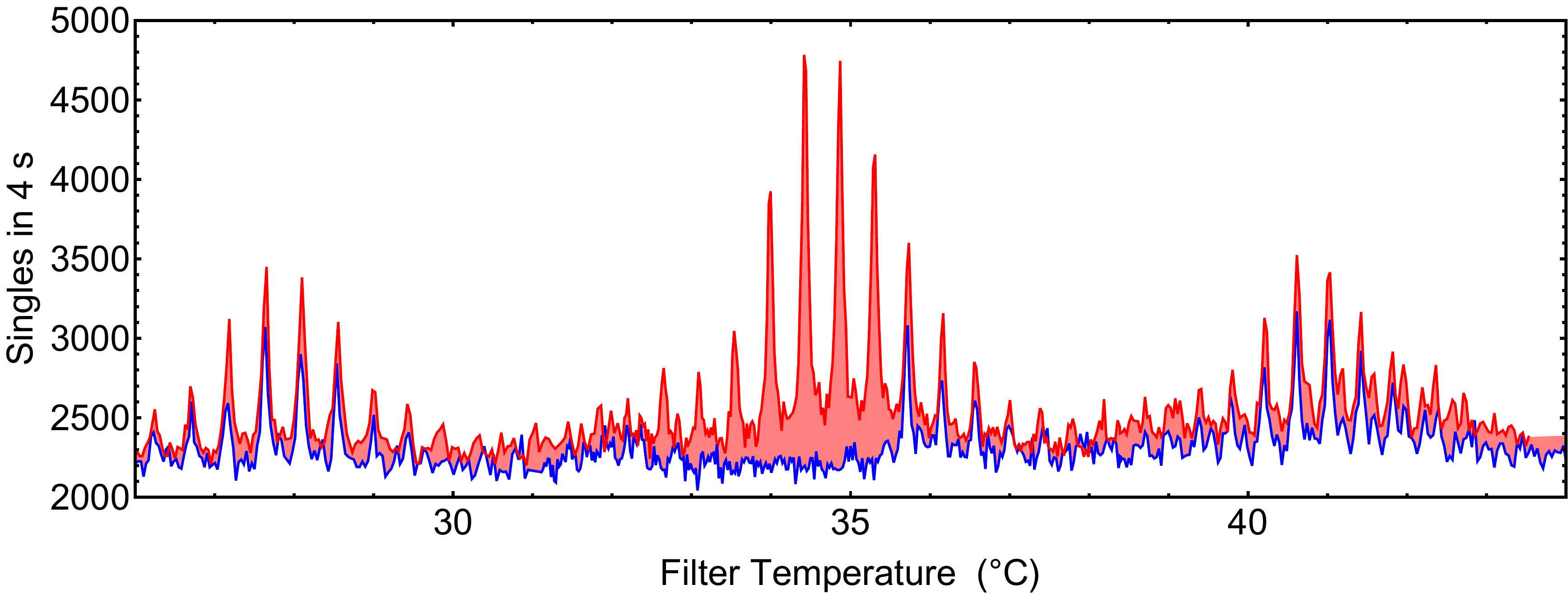} 
\caption{Singles detection of signal photons as FP filter resonance is scanned. The \CESPDC~ is pumped with \SI{4}{\milli\watt} and tuned such that, in the brightest cluster, signal photons are resonant to the $F=2$ to $F'=1$ transition of $^{87}$Rb. The signal photons are passed through the FP filter and a Rb vapor cell and  detected with an APD.  Filter temperature is scanned in steps with \SI{4}{\second} acquisition at each temperature.  Red filled region shows observed singles with the vapor cell at room temperature. Due to the \SI{39.4}{\giga\hertz} FSR of the filter cavity, the three clusters within the SPDC bandwidth are visible in this scan of width $\approx \SI{21}{\giga\hertz}$ (see text). Blue curve shows singles with the vapor cell at \SI{90}{\degreeCelsius}. The modes from the brightest cluster are blocked by atomic absorption, with the singles count dropping to the background level. In contrast, the other two clusters are unaffected.}
\label{fig:FilterScan}
\end{figure*}

To measure  $g_{s,i}^{(2)}(\tau)$, we detect signal and idler photons with avalanche photodiodes (APDs with a quantum efficiency of 49$\%$),\commentout{ Perkin Elmer SPCM-AQ4C )} and detection events are time tagged using a field programmable gate array (FPGA),\commentout{ Arty A7-35T: Artix-7, XC7A35TICSG324-1L from Xilinx programmed using the Xilinx Software Development Kit version 2018.3),} programmed as a time-to-digital converter (TDC) with \SI{625}{\pico\second} resolution. Results are shown in \autoref{fig:G2}. Fitting \autoref{eq:g2} to the data of \autoref{fig:G2}(b) gives the cavity relaxation rates $\gamma_{s}  =2\pi \,\times \SI{6.9\pm 0.2}{\mega\hertz}$ and $\gamma_i =2\pi \,\times \SI{6.3\pm 0.1}{\mega\hertz}$. The linewidth of the correlated photons is thus $\SI{4.3\pm 0.4}{\mega\hertz}$ following \cite{OBenson3}. We measure a pair brightness of \SI{180}{coincidences\per\second\per\milli\watt} and a heralding efficiency of 9$\%$. Correcting for the quantum efficiency of the APDs, we have \SI{367}{pairs\per\second\per\milli\watt} of correlated photons available in fiber for interaction with cold atoms and an 18$\%$ probability of having a photon in fiber, on detecting a herald.  When the filter is introduced the comb structure is washed out from the $g_{s,i}^{(2)}(\tau)$. This indicates the absence of JSI frequency components detuned from the majority mode by less than \SI{1.6}{\giga\hertz}  (the inverse of the \SI{625}{\pico\second} time resolution), as expected from the predicted filter behavior.

\subsection{Analysis of \CESPDC~spectral content}\label{sec:SpectralContentAnalysis}

We further analyse the spectral content of the \CESPDC~output by measuring the singles as the filter is scanned in frequency. For this measurement, the \CESPDC~is set to generate signal photons resonant to the $F=2$ to $F'=1$ transition in $^{87}$Rb. The signal photons are collected, filtered through the tuneable filter, passed through a room temperature natural abundance Rb vapour cell with internal length 10 cm , and sent to an APD for counting. The temperature and thus the resonance frequency of the filter is swept to give the spectrum of the signal photons as shown in the red filled plot in \autoref{fig:FilterScan}. 

The scan of the cavity, which covers about \SI{11}{\giga\hertz} on either side of the  $F=2$ to $F'=1$ transition, nonetheless shows the three clusters, predicted by theory to be found \SI{-70}{\giga\hertz}, \SI{0}{\giga\hertz} and \SI{+70}{\giga\hertz} relative to this atomic line; see \autoref{fig:clusters}. The clusters are aliased into the range of the scan by multiples of the 39.4 GHz FSR$_{\rm FP}$. In this way, the cluster at \SI{+70}{\giga\hertz}, e.g., is expected to appear at $70-2\times 39.4 = \SI{-8.8}{\giga\hertz}$, and indeed a cluster is seen at this detuning.

Because the phase matching was optimised for signal photons at the $F=2$ to $F'=1$ transition in $^{87}$Rb, the central mode of the brightest cluster should correspond to this frequency. To verify this, we heat the cell to \SI{90}{\degreeCelsius} and thereby induce an atomic density sufficient to completely block light resonant to the $F=2 \rightarrow F'=1$ and $2$ transitions in $^{87}$Rb, and $F=3 \rightarrow F'=2$ and 3 transitions in $^{85}$Rb (checked by measuring the transmission of coherent light of $\sim$ \SI{1}{\milli\watt}). These transition, spaced apart by 816, 702 and 361 MHz, with their individual Doppler broadened linewidths of $\approx$ 550 MHz, effectively block light within a 3 GHz window. The transmission of the other clusters remains unaffected. It is also noteworthy that the attenuated portion of the spectrum drops to the background level.  This shows that any ``junk photons'' (i.e., not from the desired cavity mode) that pass the filter must be within the spectral range blocked by the vapor.  In combination with the $g_{s,i}^{(2)}(\tau)$ results above, which show no junk photons within $\pm \SI{1.6}{\giga\hertz}$ of the desired mode, this proves single-mode emission to within the sensitivity of these measurements.

\subsection{Atomic Spectroscopy with \CESPDC~photons}

\begin{figure}[t]
\centering
\subfigure[\ ]{
\includegraphics[width=0.80\columnwidth]{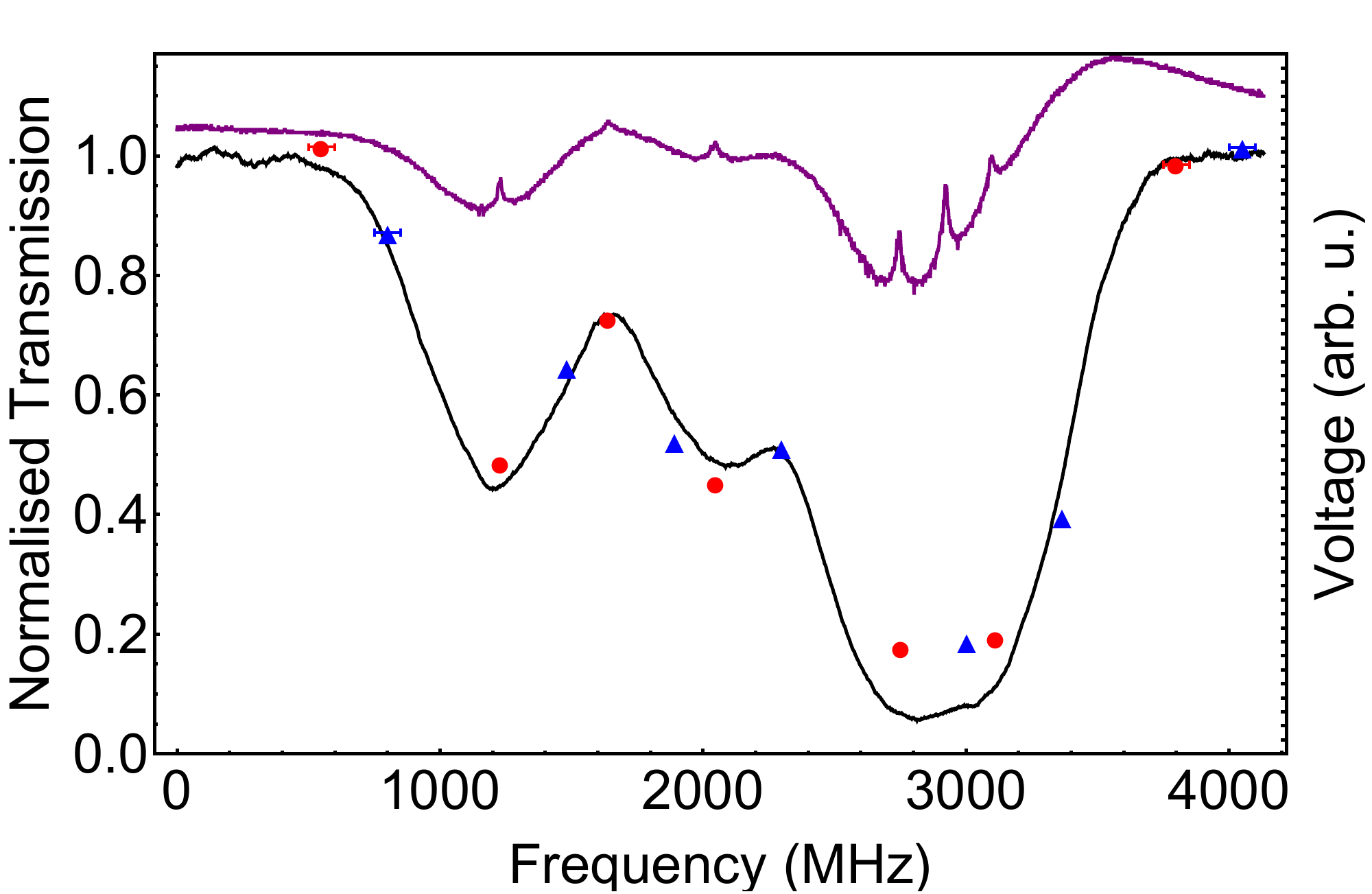}}
\subfigure[\ ]{
\includegraphics[width=0.80\columnwidth]{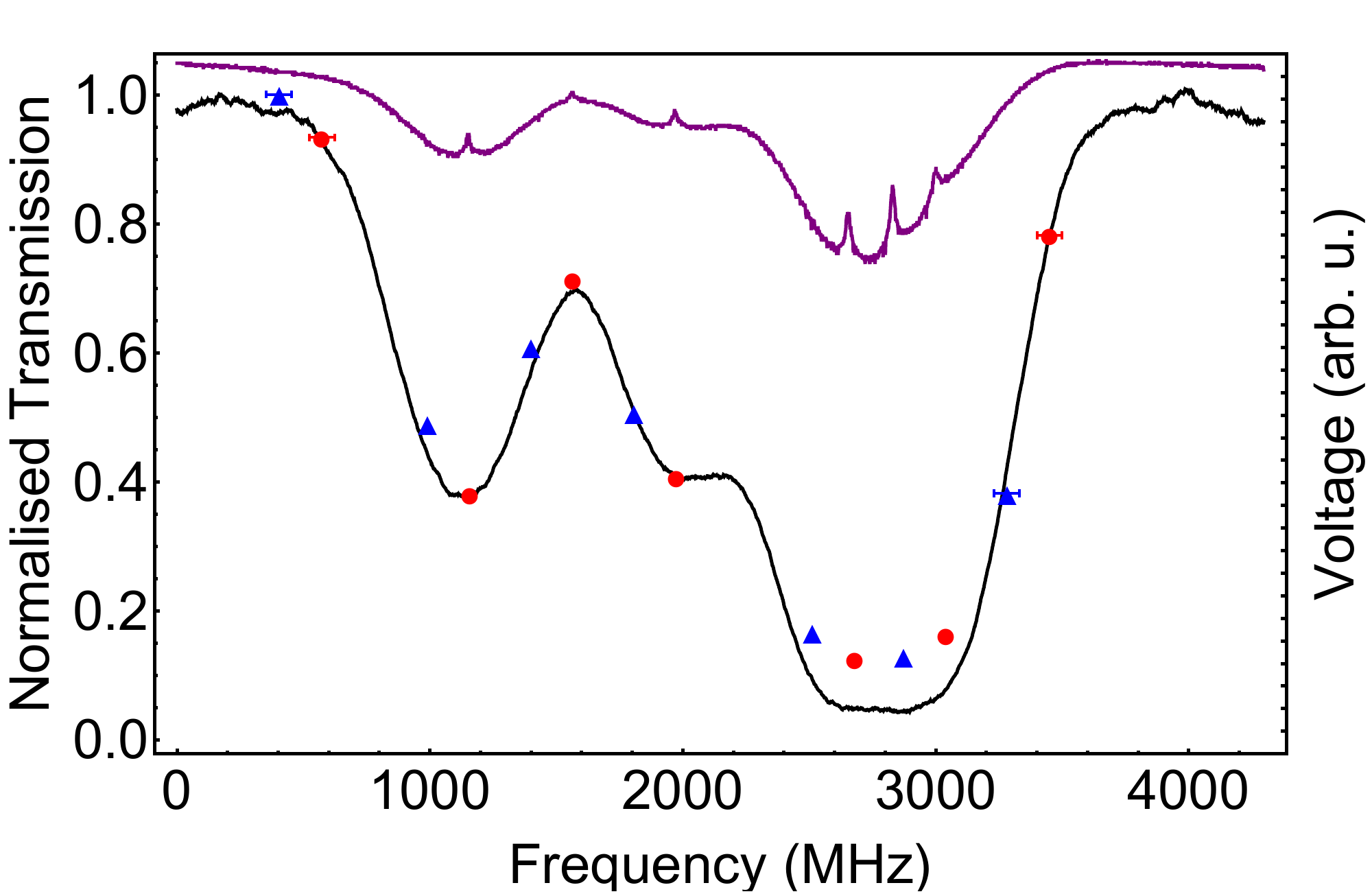}}
\caption{Atomic vapor spectra acquired with \CESPDC~photons.  Black curves show measured transmission of a weak laser through a heated, natural-abundance vapor cell. Violet curves show saturated absorption spectrum (right axes) with another cell for reference. Red disks and blue triangles show signal and idler transmission, respectively, through the heated cell. 
Each data point corresponds to an acquisition time of \SI{12}{\second} at \SI{4}{\milli\watt} pump power, which yielded roughly 20,000 detections  in transparent regions of the spectrum.  The  \CESPDC~photons were tuned as described in \autoref{sec:tunablephotons} and filtered to single-mode as described in \autoref{sec:FilterCavity}.  When possible, $\sigfreq$ was stabilized to a feature of the saturated absorption spectrum, while $\idlfreq$ was stabilized to a) $\sigfreq + \SI{250}{\mega\hertz}$ or b) $\sigfreq - \SI{170}{\mega\hertz}$. At the edges of the spectrum $\sigfreq$ was not actively stabilized, and horizontal error bars indicate the uncertainty in the estimated frequency of the lock light and consequently the signal/idler photons. Poisson distributed noise in the detected photons would contribute vertical error bars smaller than the symbols and are not shown.
}
\label{fig:TuneablePhotons}
\end{figure}

To test the independent tuneability of the \CESPDC~ signal and idler, we performed single-photon spectroscopy on the same Rb vapor cell of 10 cm length.  The cell was heated to \SI{40}{\degreeCelsius}, a temperature at which the absorption at different transitions can be clearly distinguished. For reference, a spectrum under the same conditions was taken with \SI{6}{\micro\watt} of laser light. This low power was chosen to avoid saturation of the spectrum by optical pumping. 

For each measurement, we tune the signal, idler and pump frequencies using the techniques described in \autoref{sec:tunablephotons}:  $\lockfreq$, and thus $\sigfreq$, is set to a feature of the D1 saturated absorption spectrum of either one of the rubidium isotopes.  $\idlfreq$ is set to $\sigfreq + \Delta\nu$ with $\Delta\nu = \SI{250}{\mega\hertz}$ or \SI{-170}{\mega\hertz}  by temperature tuning of the tuning crystal.  The pump is locked to $2\sigfreq + \Delta\nu$ to satisfy energy conservation.  Either signal or idler, filtered with a tunable filter (as in \autoref{sec:FilterCavity}) which is tuned to pass the corresponding frequency,  is passed through the vapour cell and detected with an APD. A beamsplitter and auxiliary APD before the cell are used for a simultaneous measurement of the source brightness.  The measured background, i.e. APD counts with the SPDC turned off by blocking the pump, is subtracted, and the singles rate normalized by the measured brightness to obtain the cell transmission.  The cell transmission thus defined is calibrated by tuning the \CESPDC~ source far from resonance, where the absorption is negligible. As shown in \autoref{fig:TuneablePhotons}, we measured the transmission for signal photons, and for their corresponding idlers, at seven different frequencies for each of the two different values of $\Delta \nu$.  The results show good quantitative agreement with the absorption spectrum as measured by a laser.  This demonstrates both the ability to generate correlated, independently-tunable photon pairs with MHz precision, and single-mode operation even as these photons are tuned over a wide range.

\section{Conclusion}
We have described a source for tuneable narrowband correlated photon pairs, each of which can be matched to the same or different transitions in the rubidium spectrum. The photon source is a type-II phase matched nonlinear crystal within a doubly-resonant optical parametric oscillator operated far below threshold. The tuneability is achieved by altering the birefringence using an additional crystal, thereby exerting control over the relative phase between the downconverted photons. A single spectral mode is selected using a tuneable filter. The operating principles behind our system can be readily extended to produce photons suitable for interaction with other atomic/molecular/ionic species, solid state quantum memories, NV centers or combinations of these without the need for frequency conversion. Because of the type-II phase matching and the freedom on the spectral properties, the photons from our source can be easily entangled in the polarisation/frequency degrees of freedom, making them suitable for quantum networking applications such as entangling different material systems. This versatile source, capable of producing both correlated and indistinguishable photons, will also open up possibilities to explore new aspects of matter-induced nonlinearity at the single photon level and exotic quantum interference effects \cite{ScaraniBS,PhotonStatistics}.

\section*{Funding Information}
This project was supported by the European Research Council (ERC) projects AQUMET (280169) and ERIDIAN (713682); European Union projects QUIC (Grant Agreement no.~641122) and FET Innovation Launchpad UVALITH (800901);~the Spanish MINECO projects MAQRO (Ref. FIS2015-68039-P), OCARINA (Grant Ref. PGC2018-097056-B-I00) and Q-CLOCKS (PCI2018-092973), the Severo Ochoa programme (SEV-2015-0522); Ag\`{e}ncia de Gesti\'{o} d'Ajuts Universitaris i de Recerca (AGAUR) project (2017-SGR-1354); Fundaci\'{o} Privada Cellex and Generalitat de Catalunya (CERCA program); Quantum Technology Flagship project MACQSIMAL (820393); Marie Sk{\l{odowska-Curie ITN ZULF-NMR (766402); EMPIR project USOQS (17FUN03); European Union's Horizon 2020 research and innovation programme under the Marie Sk\l{}odowska-Curie grant agreement No 665884.
\section*{Acknowledgments}

The authors thank Prof. Dr. Margherita Mazzera for useful insights on \CESPDC~ sources, Dr. Alvaro Cuevas for contributions to development of the system, Dr. Simon Coop for the valuable discussions on micro-controllers and Nat{\'a}lia Alves for discussions and feedback on the manuscript.


\bibliographystyle{../biblio/apsrev4-1no-url}
\bibliography{../biblio/FinalTuneablePhotonsBib}

\end{document}